\documentclass[12pt,titlepage]{article}
\usepackage{amsmath}
\usepackage{amssymb}
\usepackage{graphicx}
\usepackage{caption2}
\usepackage{amsfonts}
\usepackage{cite}
\usepackage{lineno}
\usepackage{color, soul}
\usepackage{ulem}
\usepackage{booktabs}
\usepackage{CJKutf8}

\newcommand{\be}{\begin{equation}}
\newcommand{\ee}{\end{equation}}
\newcommand{\bea}{\begin{eqnarray}}
\newcommand{\eea}{\end{eqnarray}}
\newcommand{\bef}{\begin{figure}}
\newcommand{\ef}{\end{figure}}
\newcommand{\bt}{\begin{tabular}}
\newcommand{\et}{\end{tabular}}
\newcommand{\bno}{\begin{enumerate}}
\newcommand{\eno}{\end{enumerate}}

\def\3{\ss}

\catcode`\"=\active
\def"{\accent'177}

\begin{document}

\begin{center}

{\Large Accurate predictions of chaotic motion of a free fall disk}

\vspace{0.3cm}
\begin{CJK}{UTF8}{gbsn}
Tianzhuang Xu (徐天壮)$^1$, Jing Li (李靖)$^1$ \footnote{Email address for correspondence: lijing\_@sjtu.edu.cn}, Zhihui Li (李志辉)$^{3, 4}$ 
and Shijun Liao (廖世俊)$^{1,2}$ \footnote{Email address for correspondence: sjliao@sjtu.edu.cn}
\end{CJK}
\vspace{0.3cm}

$^1$ Center of Advanced Computing, School of Naval Architecture, Ocean and Civil Engineering, Shanghai Jiaotong University,  200240 Shanghai,  China

$^2$ School of Physics and Astronomy \\ Shanghai Jiaotong University,  200240 Shanghai,  China

$^3$ National Laboratory for Computational Fluid Dynamics,  100191 Beijing,  China

$^4$ China Aerodynamics Research and Development Center, 621000 Mianyang,  China  

 \end{center}

\hspace{-0.85cm} {\bf Abstract}  {
It is important to know the accurate trajectory of a free fall object in fluid (such as a spacecraft), whose motion might be chaotic in many cases.   However, it is impossible to accurately predict  its chaotic trajectory in a long enough duration by traditional numerical algorithms in double precision.     In this paper, we give the accurate predictions of the same problem by a new strategy, namely the Clean Numerical Simulation (CNS).  Without loss of generality, a free fall disk in water is considered, whose motion is governed by the Andersen-Pesavento-Wang model.  We illustrate that convergent and reliable trajectories of a chaotic free fall disk in a long enough interval of time  can be obtained by means of the CNS, but different traditional algorithms in double precision give disparate trajectories.  Besides,  unlike the traditional algorithms in double precision, the CNS can predict the accurate posture of the free fall disk near the vicinity of the bifurcation point of some physical parameters in a long duration.  
Therefore, the CNS can provide reliable prediction of chaotic systems in a long enough interval of time.         
}

\hspace{-0.85cm} {\bf Key words}  free fall;  chaos; numerical noises;  re-entry

\hspace{-0.85cm} {\bf AMS Subject Classifications}   37D45,  34C28,  65P20, 65M12

\section{Introduction}

Free fall motion widely exists in nature and industry.
For example, falling leaves and feathers are the common phenomena for people to see in daily life.
A very important application is the spacecraft re-entry,
since some objects are massive in size and high-temperature resistant, 
and the re-entry may cause structural, environmental and safety issues on the Earth’s surface. 
Recently, the re-entry of China’s Tiangong-1 spacecraft draws much attention to accurately predicting the uncontrolled trajectories\cite{Tang2015}.
Six degrees of freedom~(DoF) orbit model is the standard model describing this problem\cite{powell1993six, zimmerman2003automated}. 
Nevertheless, accurately modeling a free-falling process is quite challenging by the 6 DoF model. 
The chaotic features inherent of this system strongly depends on initial conditions~(SDIC)\cite{aslanov2016chaotic} and, hence, is a great threat to numerical calculation. 

Unfortunately, traditional numerical methods are naturally not ``clean''.
More or less noises (i.e., truncation errors and round-off errors) are yielded during calculation. 
These noises will increase exponentially in chaotic cases, due to the SDIC, which was first discovered by Poincaré\cite{poincare1890probleme} in 1890 and developed by Lorenz\cite{lorenz1963deterministic} in 1963, i.e., the so-called ``butterfly effect''. 
Thus, for a chaotic dynamic system, a tiny variation of the initial condition can result in significant differences between numerical trajectories after a long time simulation\cite{sprott2010elegant, sprott2003chaos}.  Furthermore,   Lorenz \cite{lorenz2006computational} reported that it is also sensitive to numerical algorithm. 
He found that the (maximum) Lyapunov exponent alters between negative and positive values even when the time step is very small.  Teixeira et al. \cite{teixeira2007time} further investigated the time step sensitivity of three non-linear atmospheric models utilizing traditional algorithms in double precision. They made a somewhat pessimistic conclusion that ``for chaotic systems, numerical convergence cannot be guaranteed forever.''

In fact, the free fall problem itself has been experiencing a long and complicated history until a widely accepted theory emerges. The qualitative discussion of  it can date back to Maxwell\cite{maxwell1854particular}. At that time, little was known about the nature of the transitions between different modes.  {Many simple shapes, such as disks, cylinders, polygons ,cones and even particles have been studied via experiments or numerical simulations of Navier-Stokes equations, and complex falling modes were discovered, including tumbling, fluttering, steady, and chaotic postures}\cite{Auguste2013, Chrust2013, Toupoint2019, Amin2019, kim2020free, esteban2018edge, esteban2019three}. Among these shapes, the disk is the most well-studied subtopic in this field. A few analytical models developed by researchers greatly enhanced the understanding of free fall motion in fluids and promised the possibility to predict the falling without solving the Navier-Stokes equations which are notorious for the high computational cost. Kuznetsov\cite{Kuznetsov2015} organized a comprehensive summary on that topic. 

A direct way to reveal the rules is the experiment.
As for the free fall motion of disks, it mainly focuses on the relationships between physical parameters and falling modes. Willmarth et al. designed a series of experiments and measured a phase diagram of fluttering, tumbling and steady descent according to six related physical parameters\cite{willmarth1964steady}. By dimensional analysis, three similarities were obtained, and, with small ratio between the thickness and the diameter of the disk, falling modes only depend on the dimensionless moment of inertia and the Reynolds number. Field et al. further found a chaotic transition region between fluttering and tumbling\cite{Field1997}. 
  {Zhong et al. recently conducted the most comprehensive experimental research on free-fall thin disks. They closely studied the relationships between fluttering (zigzag in their words) free-fall motion and Reynolds number. They found a critical Reynolds number $Re_{cr} \approx 2000$. The oscillatory amplitude is proportional to $Re$ below $Re_{cr}$ while invariant beyond $Re_{cr}$} \cite{zhong2013experimental}. 
  {Zhong et al also researched the mechanism how two-dimensional fluttering modes transform to three-dimensional spiral modes}\cite{zhong2011experimental, lee2013experimental}.   {Their experimental results have been numerically confirmed with immersed boundary-lattice Boltzmann flux solver}\cite{wang2016efficient, wang2016numerical}.   {Based on the three-dimensional spiral modes, Kim et al. further studied the free-fall motion of a pair of rigidly linked disks. They discovered a mutative falling mode with two disks falling in helical and conical motions}\cite{kim2018free}.   {Recently, Lee et al also studied the bristled disk where they found that the bristled structure of disks could strengthen the stability of free-fall motion}\cite{lee2020stabilized}.

Naturally, analytical models were built to explain the various modes and the bifurcation.
Kirchhoff made a pioneering distribution by deriving finite-dimensional governing equations\cite{kirchhoff1870ueber}, called Kirchhoff equations, based on an important fact that velocities of the solid body moving in an ideal incompressible fluid can be decoupled from the field equations of the fluid itself. 
Nevertheless, Kirchhoff equations are only related to ideal fluid corresponding to conservative systems without considering dissipation. 
It can describe the steady fall regimes and to sustain regular or chaotic oscillations and rotations\cite{borisov2006motion, borisov2007asymptotic}, but still far away from the real motion.
Inherited from Kirchhoff equations, many researchers tried to introduce appropriate amendments to build a better model, especially on replicating the fluttering, tumbling, stable and chaotic falling modes and also manifesting the bifurcation structure between fluttering and tumbling.
There are two models with significant importance, and we introduce them briefly.  
The first is the Tanabe–Kaneko model\cite{Tanabe1994}. This model firstly introduces the Joukouvsky theorem to introduce the effect of circulation and implements a sign function to relate the lift term with the kinematic information of a disk. 
Tanabe and Kaneko explained that the introduction of Joukouvsky theorem and expressed the circulation with a sign function may give rise to complex dynamics and chaos during falling in a fluid due to gravity. 
Though there was some incorrectness of the Tanabe–Kaneko formulation, including that they omitted the effect of added mass and Archimedean buoyancy and there was some contradiction between the coefficients, which was criticized after publication\cite{mahadevan1995comment, tanabe1995tanabe}, the Tanabe-Kaneko model qualitatively gives a reasonable picture of possible regimes of complex dynamics for a disk falling in a fluid.  
Andersen, Pesavento, and Wang proposed a more elaborate model to describe the fall of a flat disk or a body with an elliptic profile in a fluid through a finite-dimensional model\cite{Andersen20051, Andersen20052}. The Andersen-Pesavento-Wang model considers the problems in  Tanabe–Kaneko model  and is more coherent with the experimental results and the numerical results from the direct numerical simulation of the Navier-Stokes equations.  

However, Andersen et al. paid more attention on the phenomenology of their model but lacked close research in the simulation part. 
Without reliable numerical simulation near the heteroclinic bifurcation region, they only mentioned the possibility that the chaotic transition region found in experiments could be the heteroclinic bifurcation in their model.
Moreover, it turned out in our simulation that the model cannot provide meaningful prediction in chaos by traditional numerical methods.
Thus, we aim to conduct reliable numerical simulations to closely study the chaotic cases and heteroclinic bifurcation regions.
The motivation of this paper is to implement a radical numerical strategy to empower the Andersen-Pesavento-Wang model with the ability to accurately predict trajectories in extremely sensitive cases. 

In the present paper, the impact of numerical noises are eliminated by a novel approach.
Liao \cite{Liao2009} suggested a numerical strategy in 2009, namely the ``Clean Numerical Simulation'' (CNS) \cite{Liao2013, Liao2014}, to overcome the limitations mentioned above of traditional algorithms in double precision. 
Employing the CNS, reliable/convergent numerical simulations of chaotic dynamical systems can be obtained in a controllable interval of time $0 \leq t \leq T_c$, where $T_c$ is called the ``critical predictable time''. 
 {Compared with the traditional validated numerics methods like interval arithmetics}\cite{tucker2011validated},  {the CNS is a practical numerical strategy. The implementation of MP makes it easier to use and computationally cheaper than interval arithmetic while the convergence checks to determine $T_c$ still practically ensure the reliability of computational results.}
This method has been proved effective to calculate reliable trajectories in many chaotic systems, including Lorenz equation\cite{Liao2009}, three-body system\cite{li2017more,li2018over,li2019collisionless}, and also spatio-temporal chaos\cite{Lin2017,hu2020risks}. 
For example, by implementing the CNS, Li et al. successfully found more than 2000 new periodic orbits of the three-body problem which was pointed out by Poincaré\cite{poincare1890probleme} as a classic chaotic system.
Most of these periodic solutions are inaccessible by traditional means\cite{li2017more,li2018over,li2019collisionless},
which illustrates the usefulness of the CNS as a powerful tool for reliable investigation of chaotic systems in physics with high fidelity.
As for the spatio-temporal chaos, Lin et al. \cite{Lin2017} used the CNS to control numerical noises smaller than the micro-level thermal fluctuations, by which it was proved that the inherent micro-level thermal fluctuations are the root source of macroscopic randomness of Rayleigh-B\'ernard turbulent convection flows.
Hu et al. \cite{hu2020risks} developed a more efficient algorithm of the CNS to simulate the one-dimensional complex Ginzburg-Landau equation(CGLE).
It further exhibits that CNS method can accurately maintain both the statistical features of spatio-temporal systems and the symmetric characteristics of the solutions in which traditional numerical treatment has failed.

We organize this manuscript as follows. The Andersen-Pesavento-Wang model and the CNS strategy are briefly introduced in section 2.  In the third section, we demonstrate the sensitivity of the free fall problem to numerical noises and the advantage of the CNS method by comprehensive comparisons from both chaotic and periodic simulations. Finally, we close with discussions and concluding remarks in the last section.

\section{Mathematical model and numerical algorithm}

\subsection{Andersen-Pesavento-Wang model}

\begin{figure}[!t]
\centering
\includegraphics[scale=1]{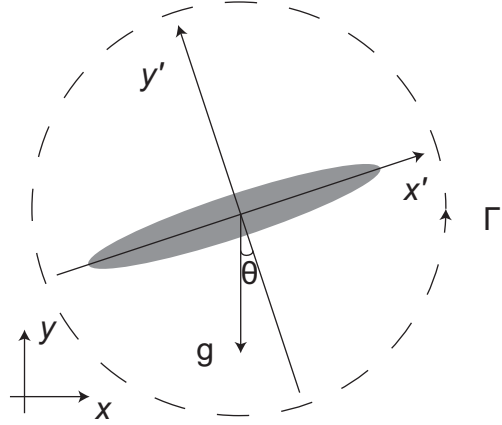}
\caption{The local reference frame $(x^{\prime}, y^{\prime})$ fixed with the disk and the global reference frame $(x,y)$ of a freely falling two-dimensional desk, where $\theta$ denotes the rotation angle of the disk, $g$ is the acceleration due to gravity,  $\Gamma$ is the circulation, respectively. }
\label{coordinate}
\end{figure}

As shown in Fig.~\ref{coordinate},  the Andersen-Pesavento-Wang model is a comprehensive analytical model to predict the trajectories of a freely falling two-dimensional disk driven by gravity: 
\begin{equation}
\left\{
\begin{aligned}
I^{*} \dot{V}_{x^{\prime}} &=\left(I^{*}+1\right) \dot{\theta} V_{y^{\prime}}-\Gamma V_{y^{\prime}}-\sin \theta-F_{x^{\prime}}, \\
\left(I^{*}+1\right) \dot{V}_{y^{\prime}} &=-I^{*} \dot{\theta} V_{x^{\prime}}+\Gamma V_{x^{\prime}}-\cos \theta-F_{y^{\prime}}, \\
\frac{1}{4}\left(I^{*}+\frac{1}{2}\right) \ddot{\theta} &=-V_{x^{\prime}} V_{y^{\prime}}-\tau,
\end{aligned}
\right.
\label{apwmodel}
\end{equation}
with the coordinate transformation
\begin{equation}
\left\{
\begin{aligned}
\dot{x}& =V_{x^{\prime}} \cos \theta-V_{y^{\prime}} \sin \theta, \\
\dot{y}& =V_{x^{\prime}} \sin \theta+V_{y^{\prime}} \cos \theta, \\
\end{aligned}
\right.
\end{equation}
where  the dot denotes the derivative with respect to the time $t$,  ($x^{\prime}, y^{\prime}$)  is the local coordinate fixed with the disk, ($x, y$) is the global (inertia) coordinate,  $\left(V_{x^{\prime}},V_{y^{\prime}}\right)$ is the velocity of disk in the local coordinate, $\theta$ is the rotation angle of the disk,  
the circulation $\Gamma$ is given by 
\begin{equation}
\Gamma=\frac{2}{\pi}\left(-\frac{C_{T}  V_{x^{\prime}} V_{y^{\prime}}}{\sqrt{V_{x^{\prime}}^{2}+V_{y^{\prime}}^{2}}}+C_{R} \dot{\theta}\right),
\end{equation}
the viscous forces $\left(F_{x^{\prime}}, F_{y^{\prime}}\right)$ and torque $\tau$ are given by
\begin{equation}
\left( F_{x^{\prime}}, F_{y^{\prime}}\right) =\frac{1}{\pi}\left(A-B \frac{V_{x^{\prime}}^{2}-V_{y^{\prime}}^{2}}{V_{x^{\prime}}^{2}+V_{y^{\prime}}^{2}}\right) \sqrt{V_{x^{\prime}}^{2}+V_{y^{\prime}}^{2}}\left(V_{x^{\prime}}, V_{y^{\prime}}\right), \hspace{0.5cm} \tau=\left( \mu_{1}+\mu_{2}|\dot{\theta}|\right) \dot{\theta},
\label{abs}
\end{equation}
respectively.  All variables are dimensionless.  This model has seven dimensionless parameters $I^*, C_T, C_R, A, B, \mu_1, \mu_2$, where $I^{*}=({\rho_{s} b}) / ({\rho_{f} a})$ with $\rho_{s}, \rho_{f}$  being the densities of disk and fluid and $a, b$  the length of the semi-major and semi-minor axis of the elliptical disk, other parameters are related with the geometry of the falling disk.  For an elliptical disk,  \cite{Andersen20051, Andersen20052}  gave  
\begin{equation} 
C_T=1.2, C_R=\pi, A=1.4, B=1.0, \mu_1=0.2, \mu_2=0.2 
\label{parameter}
\end{equation}
by means of  fitting the viscous forces and torques  of  experimental results and the direct numerical simulation of Navier-Stokes equation.  For details, please refer to  \cite{Andersen20051, Andersen20052}.  In this paper, we use the Andersen-Pesavento-Wang model  with the above-mentioned values of the parameters $C_T, C_R, A, B, \mu_1, \mu_2$.  In this paper we only adjust the parameter $I^*$ to produce different falling modes of the disk.

\subsection{Clean Numerical Simulation~(CNS)}
Generally, the CNS is able to obtain long-term reliable results thanks to its strategy to control the ``numerical noises'', say, decrease the truncation errors to a required level by implementing numerical schemes with extremely high precision and control the round-off errors within a required range with all physical/numerical variables and parameters in multiple-precision.

Truncation errors come from the discretization of continuous systems. The numerical methods have the following general form:
\begin{equation}
f(t+h) = f(t) + h \times RHS(t)
\end{equation}
where $RHS(t)$ denotes the right hand side. It varies according to the numerical methods. For N-th order Runge-Kutta family method, N-step multi-step method and N-th order Taylor series method, the right hand side have the general forms:
\begin{equation}
\left\{
\begin{aligned}
RHS(t) &= \sum_{i=1}^N k_i f(t_{k_i})+ O(h^N)\\ 
RHS(t) &= \sum_{i=1}^N k_i f(t_{i-N-1})+ O(h^N)\\
RHS(t) &= \sum_{i=1}^N \frac{f^{(i)}(t)}{i!}h^{i-1}+ O(h^N)
\end{aligned}
\right.
\end{equation}
$O(h^N)$ is the order of global truncation errors. The CNS is aimed to reduce the truncation errors so small that it would not damage the long term prediction, either by reducing time steps $h$ or increasing $N$ with high order methods. The round-off errors invariably arise with data are stored in computers in finite digits. We implemented the multiple-precision libraries (MP, called the MPFR library in the C language)\cite{oyanarte1990mp} to also reduce the round-off errors to a small enough level. 

By that, the numerical noises of the simulation are controlled arbitrarily small. To determine the critical predictable time $T_c$, one would conduct an additional simulation with even smaller numerical noises. In a temporal dynamic system, we assume that numerical noises grow exponentially within an interval of time $t \in [0, T_c]$:
\begin{equation}
\mathcal{E}(t)=\mathcal{E}_{0} \exp (\kappa t), \quad t \in\left[0, T_{c}\right]
\end{equation}
where the constant $\kappa> 0$ is called noise-growing exponent, which is coherent with the largest Lyapunov exponent~(LLE), $\mathcal{E}_{0}$. 
It denotes the level of initial noises (i.e., truncation and round-off errors), and $\mathcal{E}(t)$ is the level of evolving noises of numerical simulation. 
Theoretically, a critical level of noise $\mathcal{E}_c$ determines the critical predictable time $T_c$ by the equation:
\begin{equation}
\mathcal{E}_{c}=\mathcal{E}_{0} \exp \left(\kappa T_{c}\right)
\end{equation}
It is obvious to tell from the above equation that the smaller initial noise $\mathcal{E}_{0}$ promises a longer $T_c$. Since the true orbits are impossible to get, the CNS implements a practical method to determine  $T_c$. Let $\Phi(t)$ be a numerical simulation reliable in $t \in [0, T_c]$ with the initial noise $\mathcal{E}_{0}$, and $\Phi^{\prime}(t)$ be another simulation (with the same physical parameters and the same initial conditions) in $t \in [0, T_c]$ with the initial noise $\mathcal{E}^{\prime}_{0}$ that is several orders of magnitude smaller than $\mathcal{E}_{0}$. 
According to the hypothesis that numerical noises grow exponentially, there sure is  $T^{\prime}_c > T_c$ and $\Phi^{\prime}(t)$ in $t \in [0,T_c]$ should be much closer to the true orbit than $\Phi(t)$. Therefore, we use the $\Phi^{\prime}(t)$, a better simulation with less numerical noises, to decide the $T_c$ of $\Phi(t)$. After obtaining the $T_c$ of $\Phi(t)$, we can name safely that $\Phi(t)$ is ``clean'' numerical simulation (CNS) in $t \in [0,T_c]$. Those, as mentioned earlier,  provide us a heuristic explanation of the strategy of the CNS. The CNS also applies to non-hyperbolic chaotic systems.

We implement the strategy above to obtain the CNS results of the Andersen-Pesavento-Wang model. Given the discontinuous term in equation (\ref{abs}), we use a fixed step fourth order Runge-Kutta method with a strict time-step in multiple precision. To determine $T_c$, an additional simulation with even smaller time steps and more digits to store data in computer to guarantee that the extra simulation contains even less numerical noises.  {For the formal definition of $T_c$, we follow the form in the paper} \cite{Liao2009}.  {With formula that $\left|1-\frac{u_{1}}{u_{2}}\right|>\delta, \quad$ at $t=T_{c}$ where $\delta=1 \%$ in this paper, we determine the exact $T_c$ of the Andersen-Pesavento-Wang model}. In the following manuscript, we would like to demonstrate where and how numerical noises can harm the fidelity of trajectory prediction in the Andersen-Pesavento-Wang model.

\section{Accurate prediction given by the CNS}
\subsection{Chaotic trajectories}
\begin{figure}[!t]
	\centering
	\includegraphics[scale=.5]{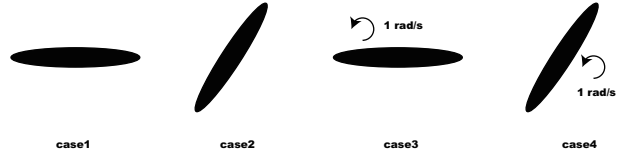}
	\caption{The schematic diagram of the four types of initial conditions considered. The four cases are correspondent to disks that fall from a static state with no angle of attack, that fall from a static state with  an angle of attack,  that fall from a rotational state with  no angle of attack and that fall from a rotational state with an angle of attack}
	\label{ic}
\end{figure}

It was regraded that chaotic free-fall motion of disks in the water is long-term unpredictable by numerical simulation. In this section, we address that the CNS is able to provide long-term reliable prediction. We study the chaotic case with $I^*=2.2$ reported by Andersen et al. \cite{Andersen20051, Andersen20052}. Four different types of initial conditions are considered for generality. These four initial conditions are listed in the form $(x, y, \theta, V_{x^{\prime}}, V_{y^{\prime}}, \dot{\theta})$ in Table.~\ref{case_table} with their corresponding schematic diagram shown in Fig.~\ref{ic}. 

\begin{table}
	\centering
	\begin{tabular}{cccc}
		\toprule  
		No. & angle of attack& rotation &$(x, y, \theta, V_{x^{\prime}}, V_{y^{\prime}}, \dot{\theta})$\\
		\midrule  
		1& & & $(0, 0, 0, 0, 0.01, 0)$\\
		2& $\surd$& & $(0, 0, 1, 0, 0.01, 0)$\\
		3& & $\surd$& $(0, 0, 0, 0, 0.01, 1)$\\
		4& $\surd$& $\surd$& $(0, 0, 1, 0, 0.01, 1)$\\
		\bottomrule 
	\end{tabular}
	\caption{The initial conditions of the four cases of chaotic falling}
	\label{case_table}
\end{table}

\begin{figure}[!t]
	\centering
	\includegraphics[scale=.6]{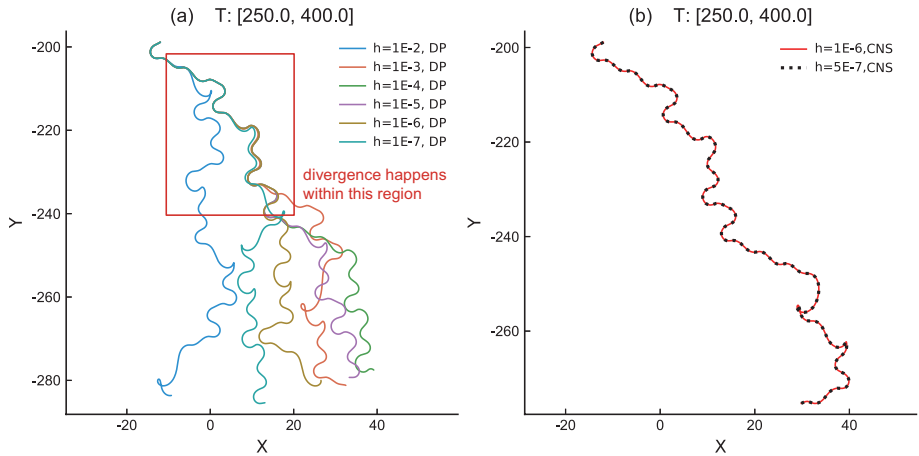}
	\caption{(a) The trajectories of the same initial condition computed by the fourth order Runge-Kutta method in double precision with different time steps. All trajectories are divergent from each other, making these trajectories of little value as prediction. (b) The trajectories of the same initial condition but computed by the CNS. The black dotted trajectory is computed with smaller time steps and data stored in more digits, which proves the red trajectory is ``clean''.}
	\label{cd}
\end{figure}

Traditional methods are powerless as a prediction tool with chaos. In Case 1, we computed trajectories by the fourth order Runge-Kutta method in double precision with different time steps from $1\mathrm{e}{-2}$ to $1\mathrm{e}{-7}$. The trajectories are plotted in Fig.~\ref{cd}, It is found that all trajectories are divergent from each other after about $250 UT$. Though traditional numerical methods can obtain qualitatively correct chaotic trajectories, these trajectories are of little use from the aspect of prediction: it's impossible to tell which one should be used as the predicted trajectory. The CNS, on the other hand, is able to give true trajectories: the trajectory with $h=1\mathrm{e}{-6}$ in 60-digits and that with $h=1\mathrm{e}{-7}$ in 120-digits. Therefore, for a chaotic system, the reliable solution is too sensitive to be found by traditional approach. The good thing is that, CNS can perfectly circumvent this issue.

\begin{figure}[!t]
\centering
\includegraphics[scale=.6]{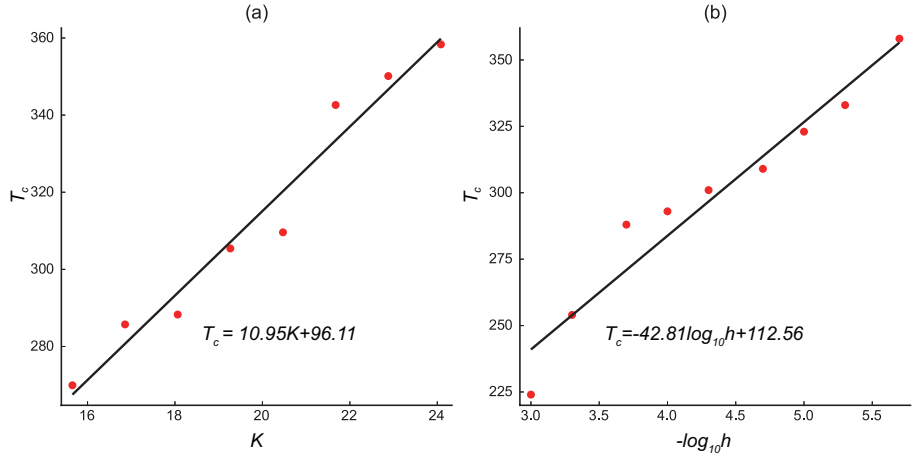}
\caption{ {(a) The linear relationships between $T_c$ and decimal digits $K$; (b) The linear relationships between $T_c$ and the logarithm of steps.}}
\label{tc}
\end{figure}

 {Without loss of generality, we consider the relationships between $T_c$ and numerical noises in case1. We have demonstrated that with $h=1\mathrm{e}{-6}$ in 60-digits the trajectory is ``clean''. According to that, two sets of numerical experiments are conducted to consider the effect of truncation and round-off errors respectively. The first set is designed to contain only truncation errors: thus all data are stored in 60-digits while the time steps are larger than $h=1\mathrm{e}{-6}$. The second set is designed to contain only round-off errors: thus the same time step $h=1\mathrm{e}{-6}$ is implemented but data are stored in less digits, from 16-digits(double precision) to around 25-digits. With the definition of $T_c$ we could compute the $T_c$ of these cases by comparing the trajectories with the ``clean'' trajectory computed with $h=1\mathrm{e}{-6}$ in 60-digits. With these two sets of numerical experiments, we confirm that $T_c$ pf the Andersen-Pesavento-Wang model also applies to the exponential growth law. There exist linear relationships between $T_c$ and the logarithm of time steps $h$, $T_c$ and the digits of data $K$, respectively. The quantitative results are given by the following equations:}
\begin{equation}
T_c \approx \min \left\{10.95K+96.11,-42.81\log_{10}{h}+112.56\right\}
\label{tcd}
\end{equation}
 {which  provide a rough estimation of the $T_c$ of case1. The exact $T_c$ is decided by the minimum of the $T_c$ with truncation and round-off errors. Generally, with smaller time steps $h$ and larger digits $K$, longer reliable prediction could be obtained. Also it is able to estimate the time steps and digits needed for a certain $T_c$ for the purpose of prediction.
}

\begin{figure}[!t]
\centering
\includegraphics[scale=.6]{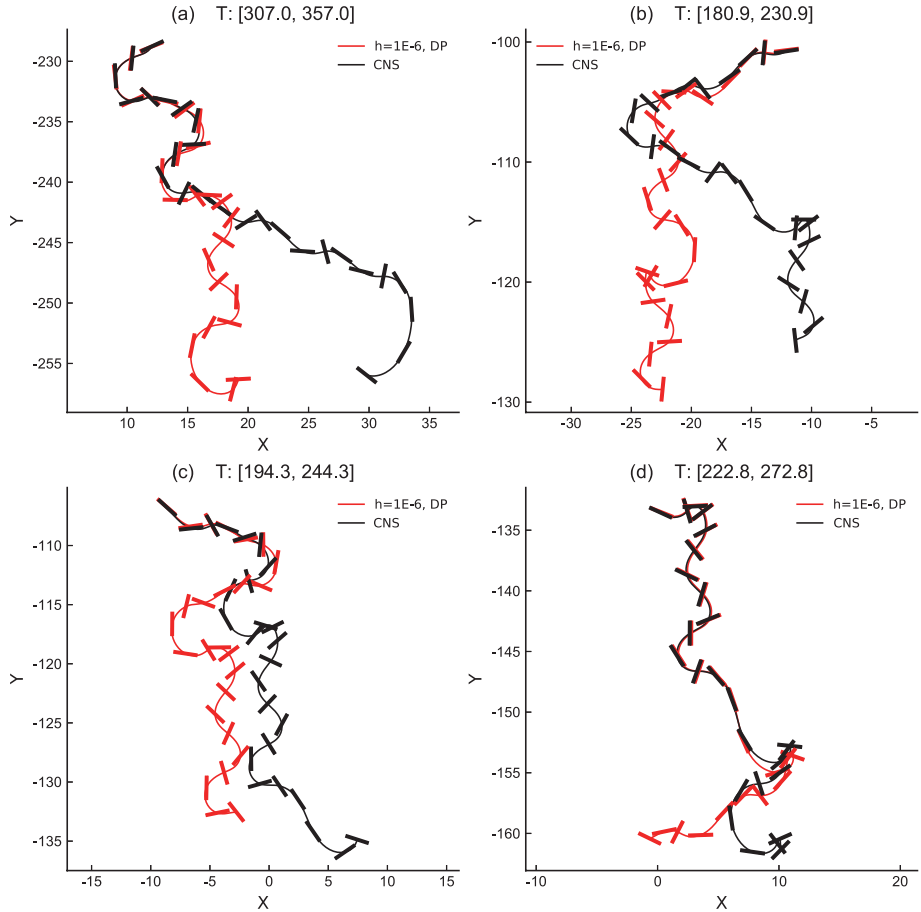}
\caption{Comparison between trajectories computed by traditional methods versus that computed by the CNS. (a) (b) (c) (d) correspond to case 1 to 4 respectively.}
\label{chaoticcases}
\end{figure}

In all four cases, the CNS is able to provide convergent prediction of the exact falling trajectories of disks. Let us compare the differences between the falling trajectories computed by the traditional methods and the CNS. The details of how trajectories computed by traditional methods diverges from the true trajectories of all four cases are demonstrated in Fig.~\ref{chaoticcases}. The trajectories in red are computed by traditional methods with $h = 1\mathrm{e}{-6}$ while the trajectories in black are computed by the CNS with the same time steps. In all our cases, the numerical noises first result in an observable phase difference of posture angle $\theta$. Then the phase difference continuously increases and finally develops into a different falling path, marking the failure of prediction. Among the four cases, the divergence happens at different times. The first case that disks fall from a static state with no angle of attack has the latest divergence, because it takes the longest time for the initial condition to develop into chaos. Except for that, these four cases have the same qualitative phenomenon.

\subsection{Heteroclinic bifurcation}

In this section, we study the heteroclinic bifurcation region reported by Andersen et al. in their paper on the analysis of transition between fluttering and tumbling state\cite{Andersen20052}. They described it as a sharp transition while the bifurcation region are sensitive to noises. In their work, Andersen et al. succeeded in studying the heteroclinic bifurcation to the precision of $1\mathrm{e}{-5}$. With the help of CNS, we approach the heteroclinic bifurcation region in even higher precision of  $1\mathrm{e}{-11}$ and provide reliable prediction which was never reported before. It is worth noting that only the CNS can predict the falling modes in that  high precision as shown in the following example.

Given the dynamical characteristics of the Andersen-Pesavento-Wang model, it has four steady solutions, corresponding to two fixed points where disk falls vertically, and gravity is balanced by drag:
\begin{equation}
\left[\begin{array}{c}
v_{x^{\prime}} \\
v_{y^{\prime}} \\
\theta \\
\dot{\theta}
\end{array}\right]=\left[\begin{array}{c}
\mp \sqrt{\frac{\pi}{A-B}} \\
0 \\
\frac{\pi}{2}, \frac{3 \pi}{2} \\
0
\end{array}\right]=\left[\begin{array}{c}
\mp V \\
0 \\
\frac{\pi}{2}, \frac{3 \pi}{2} \\
0
\end{array}\right]
\end{equation}
and two other fixed points where the face of disk is normal to the direction of motion:
\begin{equation}
\left[\begin{array}{c}
v_{x^{\prime}} \\
v_{y^{\prime}} \\
\theta \\
\dot{\theta}
\end{array}\right]=\left[\begin{array}{c}
0 \\
\mp \sqrt{\frac{\pi}{A+B}} \\
0, \pi \\
0
\end{array}\right]=\left[\begin{array}{c}
0 \\
\mp W \\
0, \pi \\
0
\end{array}\right]
\end{equation}

\begin{figure}[!t]
\centering
\includegraphics[scale=.6]{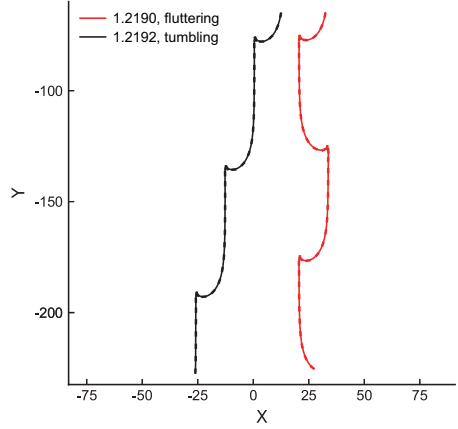}
\caption{Two characteristic trajectories near the $I^*_c$. The trajectory with $I^* = 1.2190$ fall in fluttering modes while the trajectory with $I^* = 1.2191$ fall in tumbling modes.}
\label{bifurcation}
\end{figure}

Andersen et al. discovered that the fluttering modes transform to tumbling modes via a heteroclinic bifurcation at a specific $I^*_c$.  That is with $I$ close to $I^*_c$, all disks with $I^* < I^*_c$ fall in fluttering modes while disks with $I^* > I^*_c$ fall in tumbling modes, as shown in Fig.~\ref{bifurcation}.  So mathematically, $I^*_c$ is the critical parameter that decides disks' falling modes. Noticing that the periodicity of disks grows exponentially longer as $I^*$ approaches $I^*_c$. Andersen et al. explained that the phenomenon results from heteroclinic bifurcation like the famous Silnikov’s phenomenon. 

\begin{figure}[!t]
\centering
\includegraphics[scale=.6]{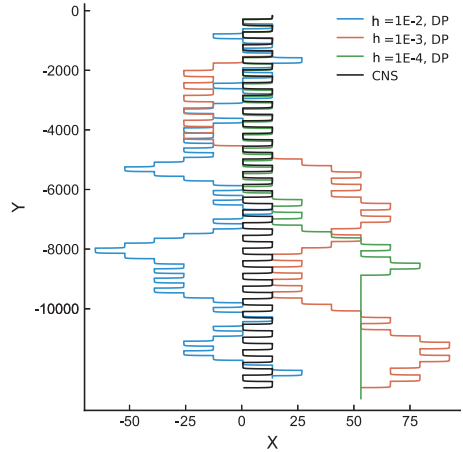}
\caption{The computational chaos resulted from numerical noises. The black line is the trajectory computed by the CNS, which falls in the fluttering modes. On the other hand, the trajectory computed by the fourth order Runge-Kutta method in double precision is neither fluttering nor tumbling. Instead they are in a special type of chaos resulted.}
\label{CC}
\end{figure}

With the help of the CNS, it is able to obtain an $I^*$ which is extremely close to $I^*_c$ by a dichotomization process based on the fact that $I^*_c$ must lie between fluttering and tumbling modes. For example, $I^* = 1.2191466312021015$ is very close to $I^*_c$ with precision  $1\mathrm{e}{-11}$. Its true trajectory is in fluttering modes computed by the CNS, plotted in Fig.~\ref{CC}. The CNS results can predict the falling modes even in extremely high-precision near the heteroclinic bifurcation point. 

However, the trajectories computed by the traditional methods with different time steps~($1\mathrm{e}{-2}$,  $1\mathrm{e}{-3}$ and $1\mathrm{e}{-4}$) behave chaotic as a Computational Chaos~(CC) phenomenon, shown in Fig.~\ref{CC}. From the aspect of prediction, that means traditional methods cannot predict the falling modes of trajectories near bifurcation point. The CC addresses a loose similarity between heteroclinic bifurcation region and chaotic transition region\cite{Field1997}. However, noting that when $I^*$ is close to $I^*_c$, the periods are elongated. It is easy to find that the ``chaotic'' trajectories in Fig.~\ref{CC} have the characteristics of periodicity, compared with the chaos discussed in the previous section in Fig.~\ref{cd}. The obvious differences confirm that the heteroclinic bifurcation region and the chaotic transition region are different.

Considering the special characteristics of the CC here, it is worth for us to distinguish it from the classic CC proposed by Lorenz\cite{Lorenz1989} as a new type. They have two main differences. First, this type of CC behaves characteristics of periodicity as mentioned. The uncertainty only happens near the heteroclinic bifurcation point that decides the disk would either flutter or tumble in the next period. Therefore, we refer such chaos as heteroclinic bifurcation chaos. Second, the CC proposed by Lorenz are only caused  by too large time steps and can be eliminated when traditional methods with smaller time steps are used. On the other hand, the CC found in the Andersen-Pesavento-Wang model is unavoidable when computed by traditional methods. No matter what time steps are picked, the CC always happens. That can be explained by that such CC can be caused both by truncation errors and round-off errors, so traditional methods is powerless to control these errors small enough to ensure the simulation ``clean''. 

Given the characteristics of heteroclinic bifurcation chaos, experimental noises, as pointed out by Andersen et al.\cite{Andersen20052}, could also trigger that heteroclinic bifurcation chaos in that region. We can only conclude the heteroclinic bifurcation of Andersen-Pesavento-Wang model is actually different from the chaotic transition region. However, we cannot preclude the existence of heteroclinic bifurcation chaos discovered here though it is strange that it was never reported before in any experiments of disks falling in the water. Whether this type of falling modes exists or the Andersen-Pesavento-Wang model is not correct as describing the transition between fluttering and tumbling is still an open question worth further study.

\section{Discussions and concluding remarks}
In this paper, we show that the uncertainty of the numerical simulation of free fall motion in the water can be overcome by a novel but powerful tool, the CNS. In the chaotic cases, numerical noises grow exponentially. Hence, even though they are as small as $1\mathrm{E}{-16}$, they are amplified exponentially.  It is impossible to obtain long-term reliable prediction of  chaotic free fall motion in the water by means of traditional numerical algorithms in double precision unless the CNS is applied.  Besides, in the heteroclinic bifurcation region, numerical noises can accumulate during the long period and destroy the bifurcation structure, behaving like a new type of computational chaos.  Under this circumstance, traditional methods cannot even predict the disk's falling modes.  Although numerical noises can ruin the validity of this problem, a good way to control them at a small enough level can still give us a trustful and reliable result in a long enough interval of time. 

In the past, the role of numerical noises was usually neglected. It is magnificent that we implement physical models to simulate and predict versatile phenomena in the nature, but it is not recommended to draw conclusions without awareness of the fidelity of the simulation.   Take the re-entry problem as an example.   In this kind of big engineering project, no doubt a better prediction of the landing spot will bring tremendous economic and safety benefits.   Hence, this work shows people a correct direction to radically solve chaotic problems by numerical methods, and it is of practical importance.

 \section*{Acknowledgments}
This work is supported by the National Natural Science Foundation of China (No. 91752104) and the project ``Development of large-scale spacecraft flight and reentry surveillance and prediction system'' of manned space engineering technology (2018-14).

 \section*{Data Availability Statement}
The data that support the findings of this study are available from the corresponding author upon reasonable request.

\bibliographystyle{elsarticle-num}
\bibliography{refs}

\begin{thebibliography}{10}
\expandafter\ifx\csname url\endcsname\relax
  \def\url#1{\texttt{#1}}\fi
\expandafter\ifx\csname urlprefix\endcsname\relax\def\urlprefix{URL }\fi
\expandafter\ifx\csname href\endcsname\relax
  \def\href#1#2{#2} \def\path#1{#1}\fi

\bibitem{Tang2015}
J.~Tang, L.~Liu, H.~Cheng, S.~Hu, J.~Duan, Long-term orbit prediction for
  tiangong-1 spacecraft using the mean atmosphere model, Advances in Space
  Research 55~(5) (2015) 1432--1444.

\bibitem{powell1993six}
R.~W. Powell, Six-degree-of-freedom guidance and control-entry analysis of the
  hl-20, Journal of Spacecraft and Rockets 30~(5) (1993) 537--542.

\bibitem{zimmerman2003automated}
C.~Zimmerman, G.~Dukeman, J.~Hanson, Automated method to compute orbital
  reentry trajectories with heating constraints, Journal of Guidance, Control,
  and Dynamics 26~(4) (2003) 523--529.

\bibitem{aslanov2016chaotic}
V.~S. Aslanov, A.~S. Ledkov, Chaotic motion of a reentry capsule during descent
  into the atmosphere, Journal of Guidance, Control, and Dynamics 39~(8) (2016)
  1834--1843.

\bibitem{poincare1890probleme}
H.~Poincar{\'e}, Sur le probl{\`e}me des trois corps et les {\'e}quations de la
  dynamique, Acta mathematica 13~(1) (1890) A3--A270.

\bibitem{lorenz1963deterministic}
E.~N. Lorenz, Deterministic nonperiodic flow, Journal of the Atmospheric
  Sciences 20~(2) (1963) 130--141.

\bibitem{sprott2010elegant}
J.~C. Sprott, Elegant chaos: algebraically simple chaotic flows, World
  Scientific, 2010.

\bibitem{sprott2003chaos}
J.~C. Sprott, J.~C. Sprott, Chaos and time-series analysis, Vol.~69, Citeseer,
  2003.

\bibitem{lorenz2006computational}
E.~N. Lorenz, Computational periodicity as observed in a simple system, Tellus
  A: Dynamic Meteorology and Oceanography 58~(5) (2006) 549--557.

\bibitem{teixeira2007time}
J.~Teixeira, C.~A. Reynolds, K.~Judd, Time step sensitivity of nonlinear
  atmospheric models: numerical convergence, truncation error growth, and
  ensemble design, Journal of the Atmospheric Sciences 64~(1) (2007) 175--189.

\bibitem{maxwell1854particular}
J.~C. Maxwell, On a particular case of the descent of a heavy body in a
  resisting medium, Camb. Dublin Math. J 9 (1854) 145--148.

\bibitem{Auguste2013}
F.~Auguste, J.~Magnaudet, D.~Fabre, {Falling styles of disks}, Journal of Fluid
  Mechanics 719 (2013) 388--405.

\bibitem{Chrust2013}
M.~Chrust, G.~Bouchet, J.~Du{\v{s}}ek, Numerical simulation of the dynamics of
  freely falling discs, Physics of Fluids 25~(4) (2013) 044102.

\bibitem{Toupoint2019}
C.~Toupoint, P.~Ern, V.~Roig, Kinematics and wake of freely falling cylinders
  at moderate reynolds numbers, Journal of Fluid Mechanics 866 (2019) 82--111.

\bibitem{Amin2019}
K.~Amin, J.~Mac~Huang, K.~J. Hu, J.~Zhang, L.~Ristroph, The role of
  shape-dependent flight stability in the origin of oriented meteorites,
  Proceedings of the National Academy of Sciences 116~(33) (2019) 16180--16185.

\bibitem{kim2020free}
J.-T. Kim, Y.~Jin, S.~Shen, A.~Dash, L.~P. Chamorro, Free fall of homogeneous
  and heterogeneous cones, Physical Review Fluids 5~(9) (2020) 093801.

\bibitem{esteban2018edge}
L.~B. Esteban, J.~Shrimpton, B.~Ganapathisubramani, Edge effects on the
  fluttering characteristics of freely falling planar particles, Physical
  Review Fluids 3~(6) (2018) 064302.

\bibitem{esteban2019three}
L.~B. Esteban, J.~Shrimpton, B.~Ganapathisubramani, Three dimensional wakes of
  freely falling planar polygons, Experiments in Fluids 60~(7) (2019) 114.

\bibitem{Kuznetsov2015}
S.~P. Kuznetsov, Plate falling in a fluid: Regular and chaotic dynamics of
  finite-dimensional models, Regular and Chaotic Dynamics 20~(3) (2015)
  345--382.

\bibitem{willmarth1964steady}
W.~W. Willmarth, N.~E. Hawk, R.~L. Harvey, Steady and unsteady motions and
  wakes of freely falling disks, The physics of Fluids 7~(2) (1964) 197--208.

\bibitem{Field1997}
S.~B. Field, M.~Klaus, M.~Moore, F.~Nori, Chaotic dynamics of falling disks,
  Nature 388~(6639) (1997) 252--254.

\bibitem{zhong2013experimental}
H.~Zhong, C.~Lee, Z.~Su, S.~Chen, M.~Zhou, J.~Wu, Experimental investigation of
  freely falling thin disks. part 1. the flow structures and reynolds number
  effects on the zigzag motion, Journal of Fluid Mechanics 716 (2013) 228.

\bibitem{zhong2011experimental}
H.~Zhong, S.~Chen, C.~Lee, Experimental study of freely falling thin disks:
  Transition from planar zigzag to spiral, Physics of Fluids 23~(1) (2011)
  011702.

\bibitem{lee2013experimental}
C.~Lee, Z.~Su, H.~Zhong, S.~Chen, M.~Zhou, J.~Wu, Experimental investigation of
  freely falling thin disks. part 2. transition of three-dimensional motion
  from zigzag to spiral, Journal of Fluid Mechanics 732 (2013) 77--104.

\bibitem{wang2016efficient}
Y.~Wang, C.~Shu, C.~Teo, L.~Yang, An efficient immersed boundary-lattice
  boltzmann flux solver for simulation of 3d incompressible flows with complex
  geometry, Computers \& Fluids 124 (2016) 54--66.

\bibitem{wang2016numerical}
Y.~Wang, C.~Shu, C.~Teo, L.~Yang, Numerical study on the freely falling plate:
  Effects of density ratio and thickness-to-length ratio, Physics of Fluids
  28~(10) (2016) 103603.

\bibitem{kim2018free}
T.~Kim, J.~Chang, D.~Kim, Free-fall dynamics of a pair of rigidly linked disks,
  Physics of Fluids 30~(3) (2018) 034104.

\bibitem{lee2020stabilized}
M.~Lee, S.~H. Lee, D.~Kim, Stabilized motion of a freely falling bristled disk,
  Physics of Fluids 32~(11) (2020) 113604.

\bibitem{kirchhoff1870ueber}
G.~Kirchhoff, Ueber die bewegung eines rotationsk{\"o}rpers in einer
  fl{\"u}ssigkeit., Journal f{\"u}r die reine und angewandte Mathematik
  1870~(71) (1870) 237--262.

\bibitem{borisov2006motion}
A.~V. Borisov, I.~S. Mamaev, On the motion of a heavy rigid body in an ideal
  fluid with circulation, Chaos: An Interdisciplinary Journal of Nonlinear
  Science 16~(1) (2006) 013118.

\bibitem{borisov2007asymptotic}
A.~V. Borisov, V.~V. Kozlov, I.~S. Mamaev, Asymptotic stability and associated
  problems of dynamics of falling rigid body, Regular and Chaotic Dynamics
  12~(5) (2007) 531--565.

\bibitem{Tanabe1994}
Y.~Tanabe, K.~Kaneko, Behavior of a falling paper, Physical Review Letters
  73~(10) (1994) 1372.

\bibitem{mahadevan1995comment}
L.~Mahadevan, H.~Aref, S.~Jones, Comment on “behavior of a falling paper”,
  Physical Review Letters 75~(7) (1995) 1420.

\bibitem{tanabe1995tanabe}
Y.~Tanabe, K.~Kaneko, Tanabe and kaneko reply, Physical Review Letters 75~(7)
  (1995) 1421.

\bibitem{Andersen20051}
A.~Andersen, U.~Pesavento, Z.~J. Wang, Analysis of transitions between
  fluttering, tumbling and steady descent of falling cards, Journal of Fluid
  Mechanics 541~(1) (2005) 91--104.

\bibitem{Andersen20052}
A.~Andersen, U.~Pesavento, Z.~J. Wang, Unsteady aerodynamics of fluttering and
  tumbling plates, Journal of Fluid Mechanics 541~(1) (2005) 65.

\bibitem{Liao2009}
S.~Liao, On the reliability of computed chaotic solutions of non-linear
  differential equations, Tellus A: Dynamic Meteorology and Oceanography 61~(4)
  (2009) 550--564.

\bibitem{Liao2013}
S.~Liao, On the numerical simulation of propagation of micro-level inherent
  uncertainty for chaotic dynamic systems, Chaos, Solitons \& Fractals 47
  (2013) 1--12.

\bibitem{Liao2014}
S.~Liao, Physical limit of prediction for chaotic motion of three-body problem,
  Communications in Nonlinear Science and Numerical Simulation 19~(3) (2014)
  601--616.

\bibitem{tucker2011validated}
W.~Tucker, Validated numerics: a short introduction to rigorous computations,
  Princeton University Press, 2011.

\bibitem{oyanarte1990mp}
P.~Oyanarte, Mp-a multiple precision package, Computer Physics Communications
  59 (1990) 345--358.

\bibitem{li2017more}
X.~Li, S.~Liao, More than six hundred new families of newtonian periodic planar
  collisionless three-body orbits, SCIENCE CHINA Physics, Mechanics \&
  Astronomy 60~(12) (2017) 129511.

\bibitem{li2018over}
X.~Li, Y.~Jing, S.~Liao, Over a thousand new periodic orbits of a planar
  three-body system with unequal masses, Publications of the Astronomical
  Society of Japan 70~(4) (2018) 64.

\bibitem{li2019collisionless}
X.~Li, S.~Liao, Collisionless periodic orbits in the free-fall three-body
  problem, New Astronomy 70 (2019) 22--26.

\bibitem{Lin2017}
Z.~Lin, L.~Wang, S.~Liao, On the origin of intrinsic randomness of
  rayleigh-b{\'e}nard turbulence, SCIENCE CHINA Physics, Mechanics \& Astronomy
  60~(1) (2017) 014712.

\bibitem{hu2020risks}
T.~Hu, S.~Liao, On the risks of using double precision in numerical simulations
  of spatio-temporal chaos, Journal of Computational Physics (2020) 109629.

\bibitem{Lorenz1989}
E.~N. Lorenz, Computational chaos-a prelude to computational instability,
  Physica D: Nonlinear Phenomena 35~(3) (1989) 299--317.

\end{thebibliography}

\end{document}